# The changing phases of extrasolar planet CoRoT-1b

Ignas A.G. Snellen, Ernst J.W. de Mooij, & Simon Albrecht

*Leiden Observatory, Leiden University, Postbus 9513, 2300 RA Leiden, The Netherlands*

**Hot Jupiters are a class of extrasolar planet that orbit their parent stars at very short distances. Due to their close proximity, they are expected to be tidally locked, which can lead to a large temperature difference between their day and nightsides. Infrared observations of eclipsing systems have yielded dayside temperatures for a number of transiting planets [1-5]. Furthermore the day-night contrast of the transiting extrasolar planet HD 189733b was mapped using infrared observations[6,7]. It is expected that the contrast between the dayside and nightside of hot Jupiters is much higher at visual wavelengths as we move shortward of the peak emission, and could be further enhanced by reflected stellar light. Here we report on the analysis of optical photometric data[8] of the transiting hot Jupiter CoRoT-1b, which cover 36 planetary orbits. The nightside hemisphere of the planet is consistent with being entirely black, with the dayside flux dominating the optical phase curve. This means that at optical wavelengths the planet's phase variation is just as we see it for the interior planets in our own solar system. The data allow only for a small fraction of reflected light, corresponding to a geometric albedo <0.20.**

The CoRoT (Convection Rotation and Planetary Transits) satellite monitored the extrasolar planet CoRoT-1b nearly continuously over 55 days, among ~12,000 other stars in its fields of view[9]. The time sampling was 512 s during the first 30 days and 32 s for the remainder of the observations, providing a light curve with nearly 69,000 data points covering 36 planetary orbital periods (P = 1.509 d). A prism in front of the exoplanet CCDs produces small spectra for each star, on which aperture photometry is performed in three bands (red, green, and blue)[8]. This is done on board to comply with the available telemetry volume. The transmission curves of the three bands are different for each targeted star. They depend on the template chosen for the on-board aperture photometry, which is based on the effective temperature of the star and its position on the CCD. The Earth has a significant influence on the photometric performance of the satellite and introduces relevant perturbations on time scales of the satellite orbital period (103 min) and the 24 hr day. Most of these effects were corrected for prior to the data release for the general astronomy community[10].

For a detailed description of the data analysis we refer to the supplementary information. Concentrating on the data from the red pass-band, we rejected outlier data points and removed residual instrumental effects on the time scale of the orbital period of the satellite and the 24-hour day. The final, corrected and partly re-sampled light curve contains 7883 data points with a relative standard deviation of $1.0 \times 10^{-3}$, with correlated noise estimated at a level of $\sim 1.2 \times 10^{-4}$, which decreases significantly by averaging the signal over 34 transits. In strong contrast to the red channel data, the light curves from the green and blue channels unfortunately exhibit ramps, sudden jumps and high levels of correlated noise, which make them unusable for the analysis carried out here. The short wavelength cut-off of the red channel pass-band was determined from the overall transmission curve for the telescope/CCD combination[8] multiplied with a Kurucz model spectrum[11] of the host star, which was compared to the fraction of photons collected in the red channel. This results in a wavelength cutoff of 560 nm, and an effective wavelength of 710 nm.

The final, corrected light curve (Fig. 1), folded over the orbital period of the planet and binned over 0.05 in phase, exhibits a distinct rise in flux over the first half of the orbit, followed by a dip at orbital phase = 0.5 and a significant decrease during the second half of the orbit. This is entirely consistent with the dayside hemisphere rotating in to view, being eclipsed by the star, and rotating out of view again. The amount of light received from the system at the moment the planet is eclipsed, is similar to that when the nightside of the planet is in full view. This means that to within the observational uncertainties we receive no light from the nightside of the planet.

We fitted the light curve with a three parameter model using a $\chi^2$-analysis, assuming a homogeneous surface brightness for each of the hemispheres. The first parameter, $R_{Day}$, is the contrast ratio between the planet dayside flux and the stellar flux. The second parameter denotes the ratio of night-side to day-side flux, $F_{N/D}$. A third parameter represents the flux of the star. We determine the two relevant parameters to be $R_{Day} = 1.26 \pm 0.33 \times 10^{-4}$ (with a null-detection rejected at $\sim 4\sigma$ level), and a best-fit value of $F_{N/D} = 0$ with an upper limit of $F_{N/D} < 0.24$ at $1\sigma$ ($< 0.47$ at $2\sigma$), meaning that the integrated light at nightside hemisphere is <24% (<47%) of that of the dayside hemisphere. The influence of correlated noise was assessed by also fitting the model to a 1 hr binned light curve, for which the error of each binned point was calculated from the variation of the points at the same orbital phase over all periods. In this way $R_{Day}$ was determined to be slightly higher at $1.47 \pm 0.40 \times 10^{-4}$, at a somewhat lower significance. In addition we performed a Markov Chain Monte Carlo (MCMC) simulation, which results in $R_{Day} = 1.40 \pm 0.33 \times 10^{-4}$ (see supplementary info). Both $R_{Day}$ and $F_{N/D}$ are influenced by the phase variation and the eclipse depth, but in different ways. While the eclipse depth is a direct measure for $R_{Day}$, and the phase variation provides a lower limit to it, it is the ratio of phase variation over the eclipse depth that governs $F_{N/D}$. The entire phase curve contributes to the estimates of $R_{Day}$ and $F_{N/D}$, including the points near the transit.

The high level of irradiation from the nearby host star is a major factor determining the atmospheric properties of a hot Jupiter. There is mounting evidence, both theoretically and observationally, that differences in incident star flux between planets leads to at least two distinct classes of hot Jupiter atmospheres, depending on whether or not their atmospheres contain highly absorbing substances such as gaseous TiO and VO[12-19]. These hot Jupiters are called "pM-" and "pL class" planets[12], analogous to the M- and L- type stellar dwarfs, and each are expected to have very different spectra and day to nightside circulations. The pM class planets, with high levels of irradiation, are thought to be warm enough to prevent condensation of titanium (Ti) and vanadium (V) bearing compounds. This leads to absorption of incident flux by TiO and VO at low pressure and subsequently a temperature inversion in the planet's stratosphere. These planets are expected to appear anomalously bright in the infrared, and to exhibit molecular bands in emission rather than in absorption[12-15]. Broadband Infrared secondary eclipse measurements for HD209458b, a planet thought to be just above the pM/pL boundary, are indeed best explained by the presence of thermal inversion and water emission bands[14,19]. In addition, the broadband infrared spectrum of the even warmer planet TrES-4b is also best fitted with models assuming a temperature inversion in its atmosphere[17]. In the atmospheres of planets which receive less stellar flux (pL class planets such as HD189733b), Ti and V are expected to be condensed out, and therefore should not exhibit thermal inversion. The recent infrared spectrum of HD189733b indeed shows strong water absorption and is best matched with models that do not include an atmospheric temperature inversion[16].

Whether or not the incident stellar flux is absorbed in a planet's stratosphere strongly influences the ratio of radiative timescales to expected dynamical timescales, and determines to what extent the absorbed energy is redistributed to the planet's nightside[12-14]. The cooler pL class planets absorb incident flux deep in the atmosphere where the atmospheric dynamics are more likely to redistribute absorbed energy, leading to cooler daysides, warmer nightsides, and strong jet-flows resulting in significant phase-shifts in their thermal emission light curves. Indeed, the pL class planet HD189733b is observed to have day/night temperature differences of ~240 K at both 8 and 24 μm, accompanied by a phase shift of 20-30°[6,7]. In pM class planets, the absorbed energy is reradiated before it could be transported to the nightside, resulting in large day/nightside temperature contrasts and neglegible phase shifts in the their thermal emission light curves. Indeed, large day/night contrasts have been found for two pM class planets, υ And b[20] and HD179949b[18], although a detailed interpretation is hampered because both systems are non-transiting and their orbital inclinations and planet radii unknown.

CoRoT-1b is a strongly irradiated planet and should therefore fall well within the pM class. The measured day/nightside flux difference is $1.26\pm0.36\times10^{-4}$, meaning that if there is no reflective component in the red channel light curve, the redistribution fraction, $P_n$, which is the fraction of absorbed stellar radiation that is transported to the night side of the planet[21], has an upper limit of $P_n<0.22$ at $2\sigma$ ($P_n<0.39$ at $3\sigma$), in line with the low distribution factors expected for pM planets. In addition there is no evidence for a phase shift in the light curve. Therefore, in the case CoRoT-1b has a low albedo, it exhibits all the characteristics expected for a pM class planet. We measure its hemisphere-averaged dayside brightness temperature to be 2390±90 K. Assuming a uniform hemispheric emission, the maximum possible brightness temperature is slightly lower at ~2260 K for a non-reflective planet with a redistribution factor of $P_n=0$. If instantaneous re-emission of absorbed radiation is assumed without advection, a maximum dayside brightness temperature of ~2430 K is obtained, within the $1\sigma$ uncertainty limits of the measured temperature. Since this indicates that any reflective component in the planet's light curve is probably small, it means that the planet has a very low geometric albedo in CoRoT's red channel.

Both theoretical modeling and observed upper limits indeed imply very low reflectivity for hot Jupiters. Ground-based spectroscopy that exploits the Doppler effect to separate the spectral lines of a hot Jupiter from the lines of its parent star, has yielded stringent upper limits for geometric albedos (e.g. $A_g<0.12$ at $3\sigma$ for HD 75289b)[22-24]. In addition, analysis of data from the MOST satellite has yielded an upper limit of $A_g<0.17$ at $3\sigma$ for HD 209458b[25]. In comparison, the geometric albedos of solar system gas giant planets range from 0.41 to 0.52[26]. Theoretical modeling of exoplanet atmopsheres shows that many parameters can cause the low albedo of hot Jupiters, in particular the strong absorption of the alkali metals sodium and potassium (and/or the formentioned TiO and VO absorption), and the sizes and types of condensates in the atmospheres. In the absence of clouds the low albedo could be due to atomic or molecular absorption[27-29].

The red channel light curve of CoRoT-1b can also be fitted with the albedo and redistribution factor allowed to vary freely. For this we assumed that the geometric albedo is independent of wavelength and is related to the Bond albedo by $A_g$ (λ)= 2/3 $A_B$, as for a diffusely scattering (Lambert) sphere. We find that in the context of this model the variation in planet/star contrast can be explained by a geometric albedo of 0.02 to 0.2 for the full range of possible redistribution factors (Fig. 2). However, assuming that the planet's albedo is at the low end of this range, in line with both theoretical modelling and observations of other hot Jupiters, the day/night temperature contrast of CoRoT-Exo-1b is high and the redistribution factor low, as expected for a highly irradiated planet.

This year we celebrate the 400th anniversary of the first published astronomical observations with a telescope by Galileo Galilei, which he used to observe the changing phases of Venus to show the true configuration of our solar system. Now exactly four centuries later, CoRoT observations have shown the changing phases of an extrasolar planet for the first time in optical light.

**Acknowledgements** We thank the CoRoT team for making the CoRoT data publicly available in a high quality and comprehensible way, which forms the basis of this study. The CoRoT space mission, launched on Dec 27th 2006, was developed and is operated by the CNES, with participation of the science program of ESA, ESTEC/RSSD, Austria, Belgium, Brasil, Germany, and Spain. We thank Rudolf Le Poole for useful discussions.

**Author information** Reprints and permissions information is available at www.nature.com/reprints. The authors declare no competing financial interests. Correspondence and requests for materials should be addressed to I.S. (e-mail: snellen@strw.leidenuniv.nl).

Figure 1. **Optical phase variation for CoRoT-1b centered on the planetary eclipse**. Background-subtracted photometry is shown from 55 days of CoRoT monitoring in its red channel, after the rejection of >3 sigma outliers, and corrections for perturbations on a 103 min orbital time scale of the satellite, and the 24 hr day. Panel **a** and **b** show the same data, phase-folded over the P = 1.5089557d planetary orbital period, but in panel **b** the data are binned by 0.05 in phase (showing the 1$\sigma$ error bars as determined from the scatter in the individual data points), and the scale of the y-axis is enlarged by a factor ~200. It is consistent with the dayside hemisphere rotating into view, being eclipsed by the star, and rotating out of view again (panel **c**). The unbinned data is fitted with a model assuming uniform surface brightnesses for each of the dayside and nightside hemispheres, which is indicated by the solid curve. The integrated flux from the day-side hemisphere relative to the stellar flux, as determined from the depth of the planet-eclipse and the phase variation, is found to be $1.26\pm0.33\times10^{-4}$. The flux from the nightside hemisphere, as determined from the difference between the secondary eclipse depth and the amplitude of the phase variation, is found to be $<3.0\times10^{-5}$ at 1$\sigma$ ($<5.9\times10^{-5}$ at 2$\sigma$) and is consistent with being entirely black. This means that the phase variation is just as we see it for the interior planets in our own solar system, as sketched in panel **c**.

Figure 2. **The optical planet/star contrast compared with models**. Panel **a** shows the planet/star contrast as determined from the phase variation and the planetary eclipse, for the dayside and nightside of CoRoT-1b (with 1$\sigma$ error bars), together with the arbitrarily scaled pass-band of the red channel data (dotted line). Panel **b** shows the allowed ranges for the geometric albedo, $A_g$, for reflected light, and for the redistribution factor, $P_n$, which indicates what fraction of absorbed stellar radiation is transported to the nightside of the planet. Here we assume wavelength-independent Lambert scattering. To obtain the planet/star contrast expected for a combination of $A_g$ and $P_n$, first the Planck curve for the planet and Kurucz model spectrum for the primary star were separately multiplied by the transmission function of the red pass-band and integrated over wavelength. These integrated fluxes were multiplied by their respective surface areas and the ratio was taken. This was subsequently added to a possible reflective component. The solid and dashed lines in panel **a** show the two most extreme cases that perfectly fit the measured planet/star contrast, for $P_n = 0.0$, $A_g = 0.08$ and $P_n = 0.5$, $A_g = 0.15$ respectively. Within the 1$\sigma$ uncertainty in the dayside planet/star contrast, the geometric albedo is constrained to within the range $0.02 < A_g < 0.20$. Assuming there is no reflected light component, the 3$\sigma$-limit to $P_n$ is <0.39.

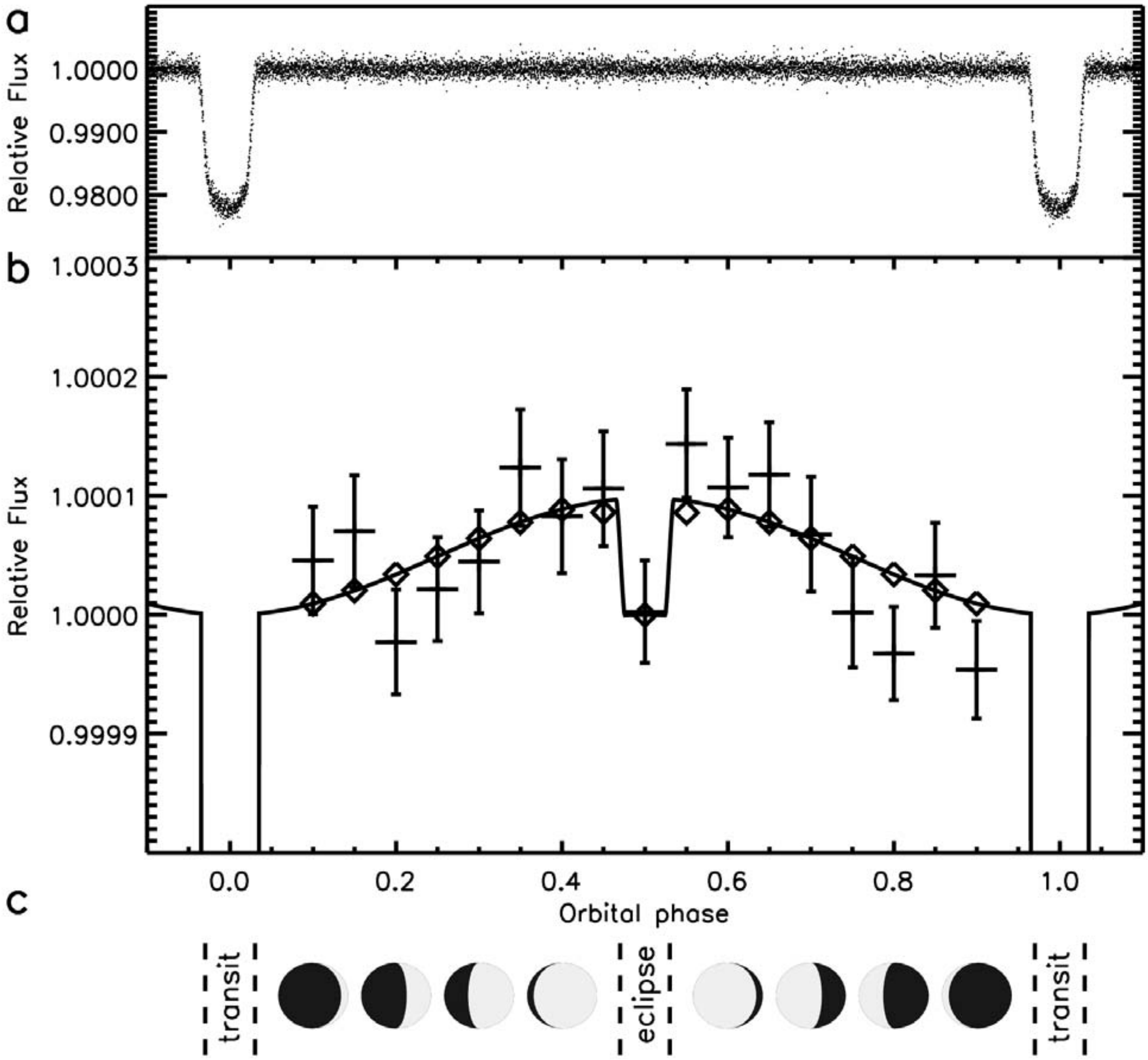
a
Relative Flux
1.0000
0.9900
0.9800
b
1.0003
1.0002
1.0001
1.0000
0.9999
Relative Flux
0.0
0.2
0.4
0.6
0.8
1.0
Orbital phase
c
transit
eclipse
transit

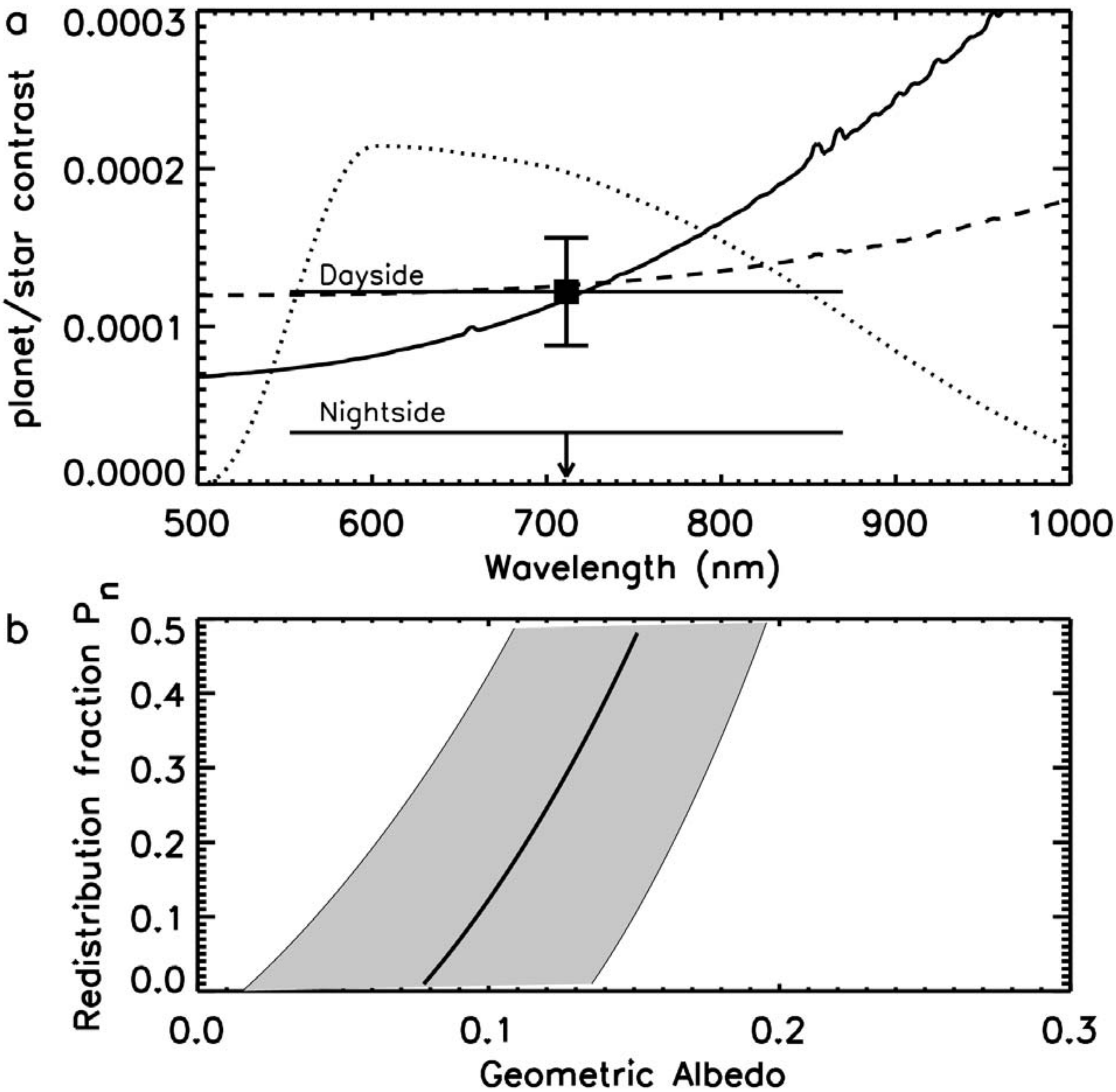
a
planet/star contrast
0.0003
0.0002
0.0001
0.0000
Dayside
Nightside
500
600
700
800
900
1000
Wavelength (nm)
b
Redistribution fraction $P_n$
0.5
0.4
0.3
0.2
0.1
0.0
0.0
0.1
0.2
0.3
Geometric Albedo

# Supplementary Information

*Ignas A.G. Snellen, Ernst J.W. de Mooij, & Simon Albrecht*

### Analysis of the red channel CoRoT-1b data

We used the updated N2 pipeline data[30] for our analysis. These were released to the general public on December 19, 2008. The satellite monitored CoRoT-1b nearly continuously over 55 days, with a time sampling of 512 s during the first 30 days and 32 s for the remainder of the observations (including ~0.4 s readout time). This provides a light curve with nearly 69,000 data points covering 36 planetary orbital periods (P = 1.509 d). A prism in front of the exoplanet-CCDs produces a small spectrum for each star, on which aperture photometry is performed in three bands (red, green, and blue[8]). The photometry is done on board to comply with the available telemetry volume. The transmission curves of the three bands are different for each targeted star, and depend on the template chosen for the on-board aperture photometry. This template is based on the effective temperature of the star and its position on the CCD. For CoRoT-1b Template ID number 174 was used, with the red channel covered by 28 pixels. The background was monitored in 400 windows of 10x10 pixels in the exoplanet field, of which 300 windows had a sampling of 512 s, and 100 windows had a sampling of 32 s[8]. The Earth has a significant influence on the photometric performance of the satellite and introduces relevant perturbations on time scales of the satellite orbital period (103 min) and the 24 hr day. These perturbations are due to ingress and egress of the spacecraft from the Earth's shadow, variations in the gravity field and magnetic field (such as the South Atlantic Anomaly), and changes in the levels of thermal and reflected light from the Earth. Most of these effects were removed before the data was released to the general astronomy community[8]. The star and planet parameters we used for our analysis are listed in Table 1.

Concentrating on the data from the red pass-band, we first divided the background-corrected light curve of CoRoT-1b in two parts, based on the sampling rate. Outliers were removed by rejecting all data points that lay more than $3\sigma$ from the median-smoothed light curve. About 85% of the rejected outliers occur during passage of the South Atlantic Anomaly or at the moment the satellite enters or leaves the Earth's shadow. Outside these periods, only 1.3% of the data points were removed. Note that changing the cut-off from $3\sigma$ to $5\sigma$ or $6\sigma$ does not alter the results. Furthermore, only a small fraction of these outliers are picked up by registered energetic particle hits and/or temperature swings. The partially observed planetary orbits at the beginning and end of the 55 day run, and during the switch between the 512 and 32 s sampling, were excluded from further analysis. This avoided introducing sudden artificial jumps in the final phased light curve. 93% of the data covering 34 full orbital periods were used. The data were phase-folded in blocks of two planetary orbital periods over 103 minutes and boxcar-smoothed over 2.5 min. The resultant curve was used to remove residual instrumental effects on the time scale of the orbital period of the satellite. An additional perturbation on a 24 hour time scale was found and removed by fitting a sinusoidal function with a 24 hour period to the entire light curve. A small part of the light curve covering one planet orbit is shown in Figure 1, illustrating the 103 min. perturbation due to the satellite orbital period (amplitude of ~0.006) and the corrected light curve. The variation with an amplitude of ~0.0007 and a 24 hour period is shown in Figure 2. The Fourier spectra of both the uncorrected and the final light curves are shown in Figure 3, indicating that the two periodicities have been largely removed. Since the light curve is re-normalised every two planet orbital periods, any variability on time scales >3 days is also removed from the data. Subsequently, the data sampled at a rate of 32 s were 16-point averaged to match the 512 s sampled data points.

The median red-channel flux of the star is ~2,340,000 $e^-$ per 512 s exposure, with an average background level of 176,000 $e^-$, resulting in a photon noise limit of 0.00068. The final, corrected and partly re-sampled light curve contains 7883 data points with a dispersion of $1.0 \times 10^{-3}$, ~50% higher than the photon noise limit. The correlated noise level in the light curve was estimated by measuring the dispersion in increasingly binned data. After comparison with the expected noise level from the unbinned data, correlated noise was estimated to be present at a level of $\sim 1.2 \times 10^{-4}$. In a previous analysis of the CoRoT-1b light curve[9], it was found that for the white-light data binned to 0.001 in phase, the noise level was a factor 3 above the photon noise limit. In our analysis the noise level is only 70% above the expected value for similarly binned data. We believe this is due to the improved N2 data pipeline.

In Figure 4 we show the uncorrected light curve, binned over 4x103 min, illustrating fluctuations on a time scale of typically ~10 days with a peak-to-peak amplitude of ~0.001. This time scale matches the expected rotational period of the primary star, for which $v\sin i$=5.2±1.0 km/s and R = 1.11±0.05 $R_{sun}$, as determined previously[9], correspond to a spin period of 10.7±2.2 days. If this interpretation is correct, the observed variability is likely to arise from star spots rotating in and out of view. There were previous indications of low frequency (< 1 cycle/day) residuals[9], and it is possible that those are responsible for some of the structure seen in Figure 4. The effect of long term variability in data from the red channel was largely removed by renormalizing the data every 2 orbital periods.

Unfortunately, data from neither the green or the blue channels can be used for the analysis described in this paper. The mean green channel flux, which is collected over only 7 pixels, is ~461,000 $e^-$ per 512 s exposure, which corresponds to a photon noise limit of 0.0015. The mean blue channel flux, collected over 26 pixels, is 755,000 $e^-$, resulting in a photon noise limit of 0.0013. Unfortunately, in contrast to the red-channel data, the blue and green data exhibit a number of jumps at the 0.5-2% level (possibly due to cosmic ray hits), in addition to numerous fluctuations on a fraction of a percent level throughout the 55 day run. This indicates that neither the blue- or green-channel data are suitable for planet phase-curve analyses.

## Uncertainties in the light curve cleaning methods

Because we removed two perturbing signals from the light curve, one on the time scale of the satellite orbit (103 min), and one on a 24 hour time scale, it is important to determine the influence of uncertainties in our cleaning methods on the final, cleaned light curve. Note that the total light curve covers ~770 satellite orbital periods and ~55 24-hour periods. Although the amplitudes of the two signals are significantly larger than the signal from the planet, the time scales are such that they have only a very small impact on the final light curve. One full period of the 103 min. perturbing signal covers only ~0.05 in planet orbital phase, which therefore effectively averages out on scales equal or larger than this. In addition, the 24 hour signal is in near-resonance with the planet orbital period (P=1.5089d), with two planet periods almost fitting exactly within 3 days. Since this perturbing signal has a sinusoidal shape, the signals at two similar points in phase, during two consecutive planet orbital periods, largely cancel each other out, because they are separated by nearly 24+12 hours.

The total effect of the two perturbing signals on the final, binned light curve is shown by the dashed line in Figure 5. It shows a peak-to-peak amplitude of ~8x10$^{-5}$, and a dispersion of 2.8x10$^{-5}$. The uncertainty in the two perturbing signals was estimated using a $\chi^2$-analysis, for which the amplitude of the signals was varied with respect to the values used above. In this way, the uncertainty in the amplitude of the 103 min. signal was determined to be ~4%, and that of the 24-hour signal to be ~12%. Assuming that both signals are uncertain by these levels, the correction to the binned light curve is shown by the solid line in Figure 5, exhibiting a peak-to-peak variation of ~4x10$^{-6}$, and a scatter of 1.2x10$^{-6}$. This shows that possible errors induced by removing the 103 min. and 24 hour perturbing signals do not significantly influence the final results.

**Error estimates of the fitted model parameters**

The light curve was fitted with a three-parameter model using a $\chi^2$-analysis, assuming a homogenous surface brightness for each of the hemispheres,

$$F(f) = [1 + \sin(\pi f)^2 \cdot (1 - P_{N/D}) \cdot R_{Day} + \psi(f) \cdot R_{Day}] \cdot z_{lev} \qquad (1)$$

where $R_{Day}$ is the contrast between the planet dayside flux and the stellar flux, $F_{N/D}$ is the ratio of night-side to day-side flux, and $z_{lev}$ is the stellar brightness. The shape of the secondary eclipse as function of phase f is denoted by $\psi(f)$, and is calculated from the planet and star parameters[9] (see Table 1) using the algorithm of Mandel and Agol[31]. The transit itself, by far the most dominant feature in the light curve, is masked off, because it has already been modelled by Barge et al[9]. The errors in the individual data points have been estimated from the rms scatter in the light curve. The analysis was performed on a grid in 3-dimensional parameter space, for which the constraints on a single parameter were determined by marginalizing over the others. In this way, we determined the two relevant parameters to be $R_{Day}$ = 1.26±0.33x10$^{-4}$ (with a null-detection rejection at ~4$\sigma$ level), and $F_{N/D}$ < 0.24 at 1$\sigma$. The influence of the correlated noise was assessed by fitting the model to a 1 hr binned light curve, for which the error of each binned point was calculated from the variation of the points at the same orbital phase over all periods. In this way, $R_{Day}$ was determined to be slightly higher at 1.47±0.40x10$^{-4}$.

As an extra check for a degeneracy between the parameters, we also performed a Markov-Chain Monte Carlo (MCMC) simulation[32], using the Metropolis-Hastings algorithm [33,34], which should result in similar uncertainties in the model parameters as above, since the full 3-dimensional parameter space was already mapped. Five independent chains were created, each from a random initial position, consisting of 10,000 points, for which the first 10% of each chain were discarded to minimize the effect of the initial condition. The Gelman & Rubin statistic[35] was calculated for each parameter to check the convergence and the consistency between the chains, which is a comparison between the intra-chain and inter-chain variance. The results were within 1% of unity, a sign of good mixing and convergence. The estimated probability distribution of $R_{Day}$ is shown in Figure 6, with the median of the distribution indicated by a solid line. The dashed lines enclose 68% of the results, with equal probability on either side of the median. This corresponds to $R_{Day}$ = 1.40±0.33x10$^{-4}$, a value in between that determined for the binned and unbinned data using the full-grid analysis. The probability distribution of $F_{N/D}$ is shown in Figure 7, with the dashed line showing the 68% upper limit at <0.26, very similar to that determined with the full-grid $\chi^2$-analysis. Figure 8 shows the same for the stellar brightness, with $z_{lev}$ expressed as a fraction of the mean flux level in the light curve ($z_{lev}$ = 0.999922±2.2x$10^{-5}$). Figures 9, 10, and 11 display $R_{Day}$ versus $P_{N/D}$, $z_{lev}$ versus $R_{Day}$, and $z_{lev}$ versus $P_{N/D}$ respectively, for the MCMC simulation, each showing a randomly selected subsample of 2x$10^3$ points for plotting purposes.

**Table 1**: The star and planet parameters from Barge et al. (2008)[9], which are used in our analysis. Parameter P is the orbital period, $T_c$ is the transit time, $T_*$ is the effective temperature of the star, $M_*$ and $R_*$ are the stellar mass and radius, $R_p/R_*$ is the ratio of planet to star radius, e is the eccentricity, and I is the orbital inclination.

| | | |
|---|---|---|
| P[d] = | 1.5089557 | ± 0.0000064 |
| $T_c$[d] = | 2454159.4532 | ± 0.0001 |
| $T_*$ [K] = | 5950 | ± 150 |
| $M_*$ [$M_{sun}$] = | 0.95 | ± 0.15 |
| $R_*$[$R_{sun}$] = | 1.11 | ± 0.05 |
| $R_p/R_*$ = | 0.1388 | ± 0.0021 |
| e = | 0 | (fixed) |
| I [deg] = | 85.1 | ± 0.5 |

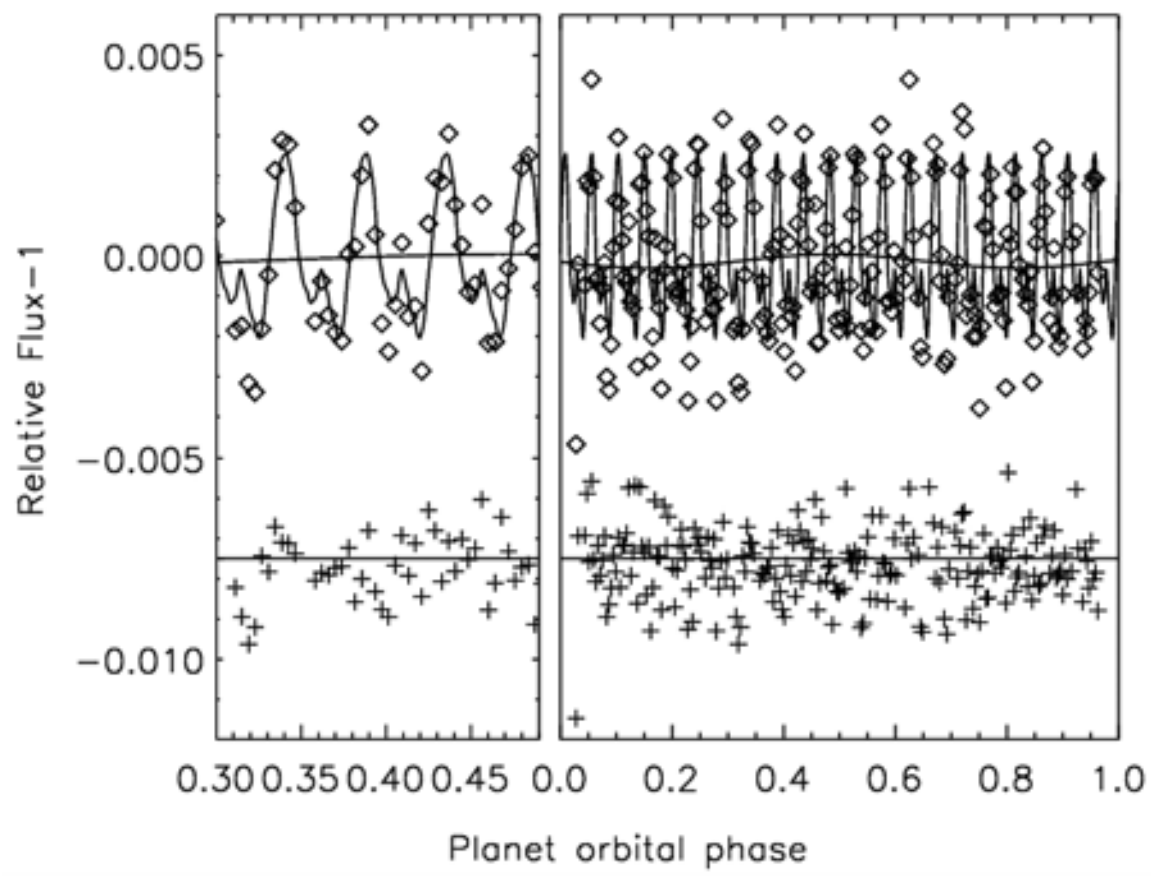

***Fig. 1:*** *The diamonds show the red channel CoRoT data, with the 3σ outliers removed, for one planet orbital period (right panel), and for a small part of this (left panel) showing the 103 min variation due to the orbital period of the satellite. The two solid lines are the corrections applied to the light curve on 103 min and 24 hour time scales. The crosses show the data after correction, offset by -0.0075.*

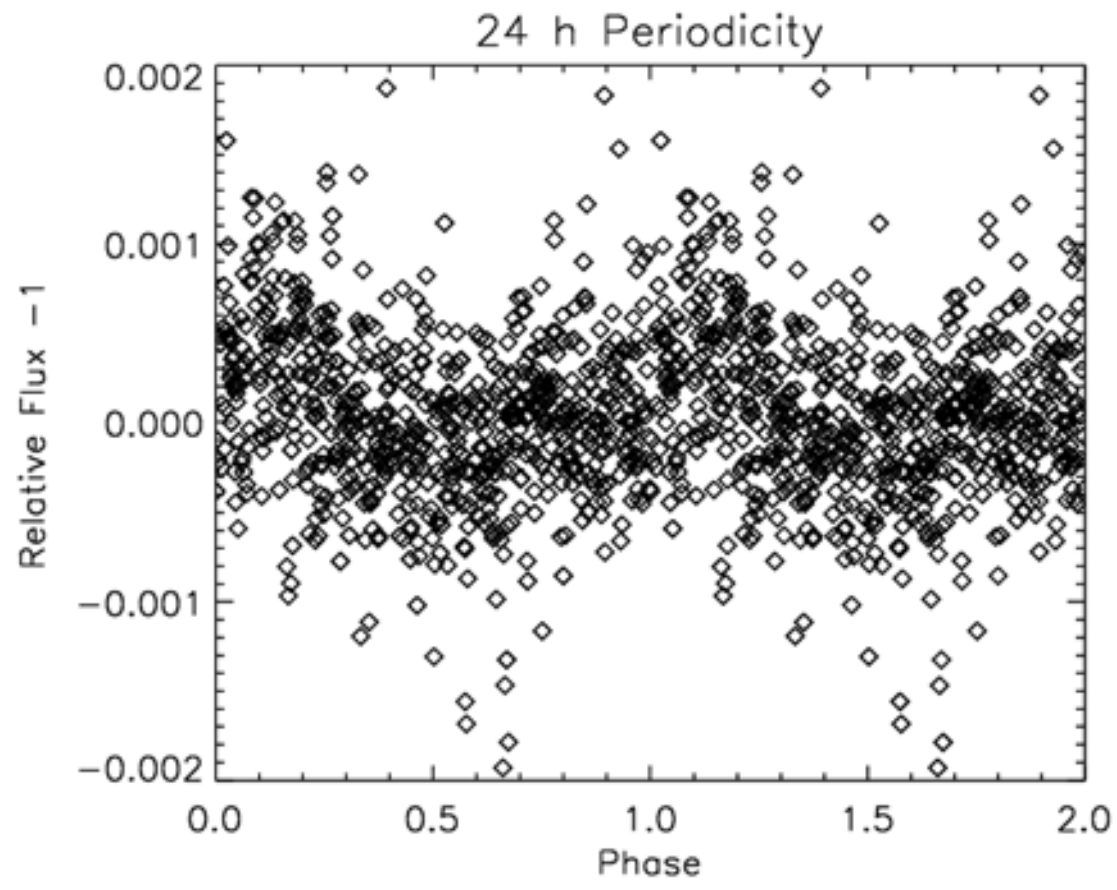

***Fig. 2:*** *The complete CoRoT light curve binned over 103 min to remove out the variations due to the satellite orbit, and subsequently phase folded over a 24 hour period. Clearly visible is the sinusoidal like variation on this time scale with an amplitude of ~0.0007.*

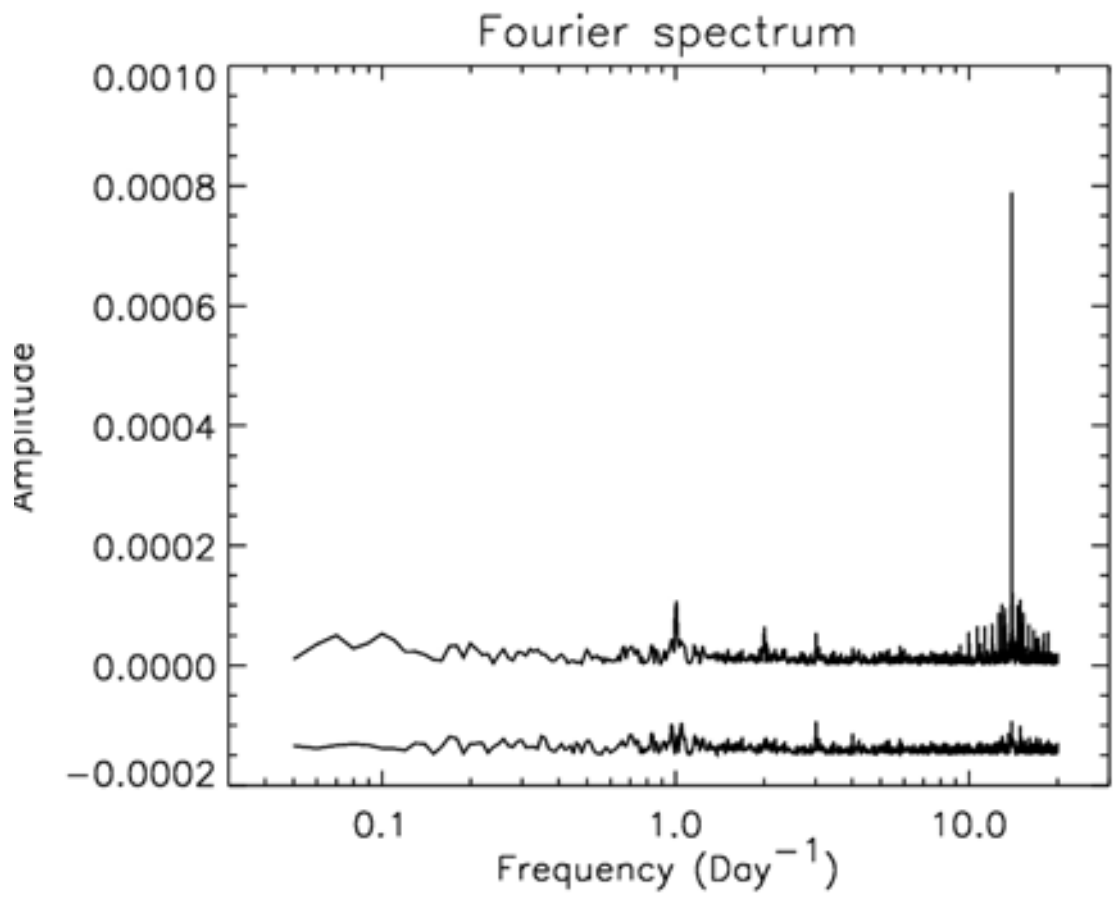

***Fig. 3:*** *The Fourier spectrum of the red channel light curve, both for the corrected and uncorrected (upper spectrum). Note that in the corrected data, the 103 min (14 cycles/day) and 24 hour periodicities are largely removed. Any variability on time scales longer than the orbital period of the planet are also mostly removed.*

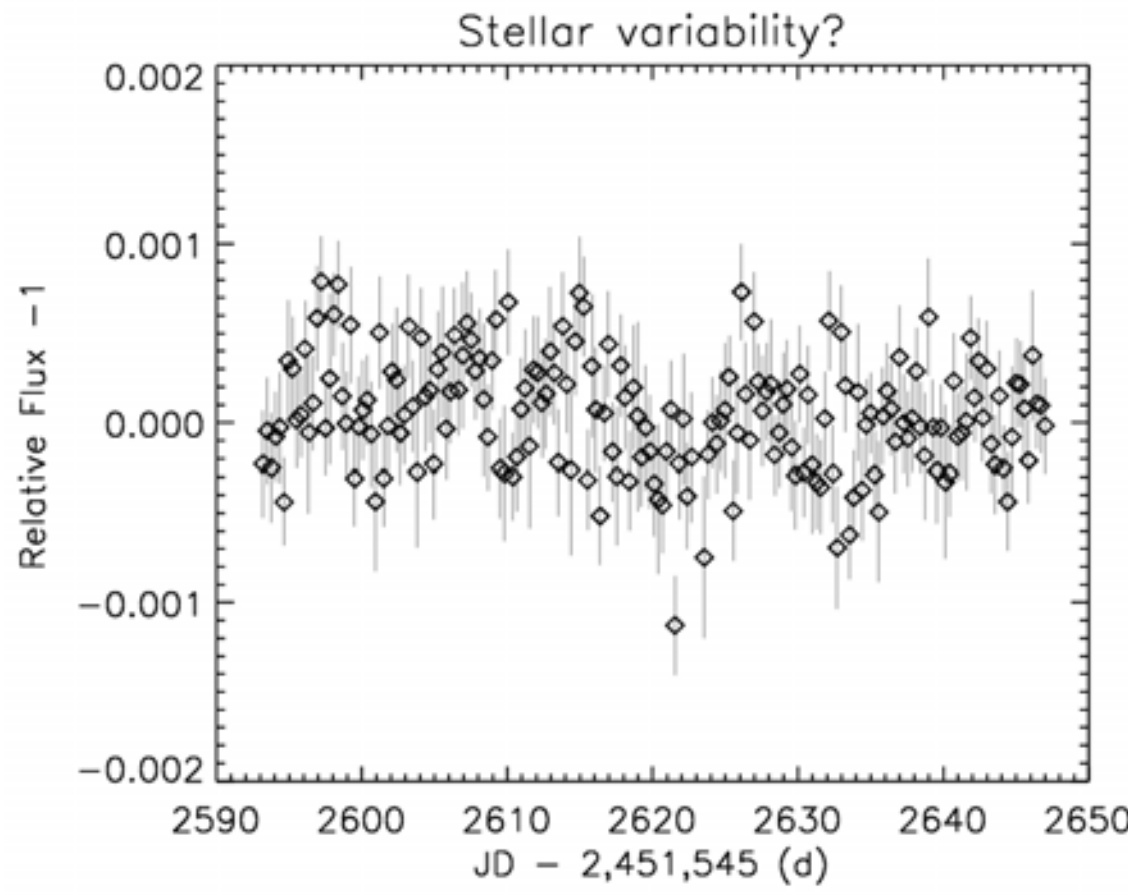

***Fig. 4:*** *The cleaned but uncorrected light curve of CoRoT-Exo-1b binned over 4x103 min(1σ error bars). There is a hint of variability on time scales of ~10 days with a peak-to-peak amplitude of ~0.001. This could well be due to spot-induced stellar variability, since the time scale corresponds to the expected rotational period of the star.*

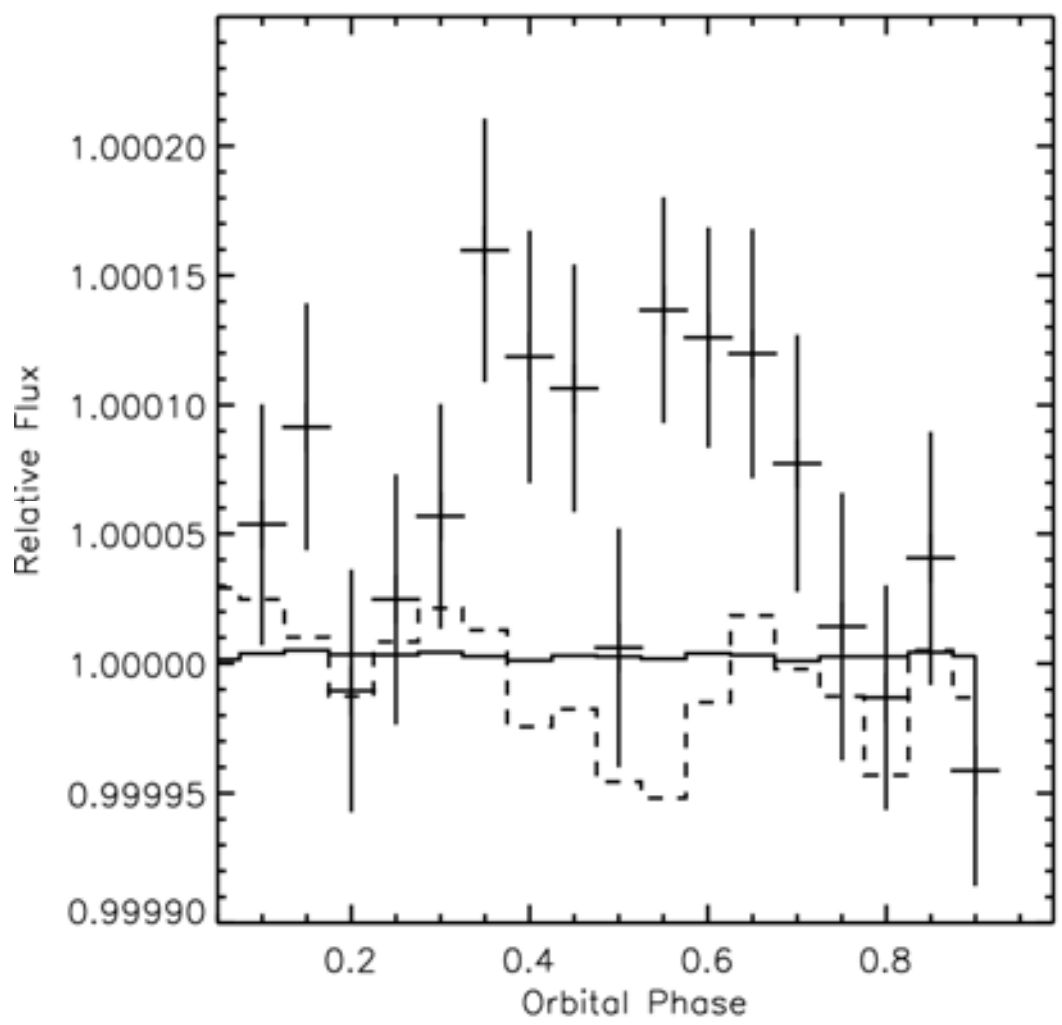

***Fig. 5***: *The corrections (dashed line) due to the 103 min. and 24 hour perturbations as applied to the final binned light curve. The solid line indicates the effect of the uncertainties in the perturbing signals on the light curve, which shows that they do not influence the final result (crosses, with 1σ error bars).*

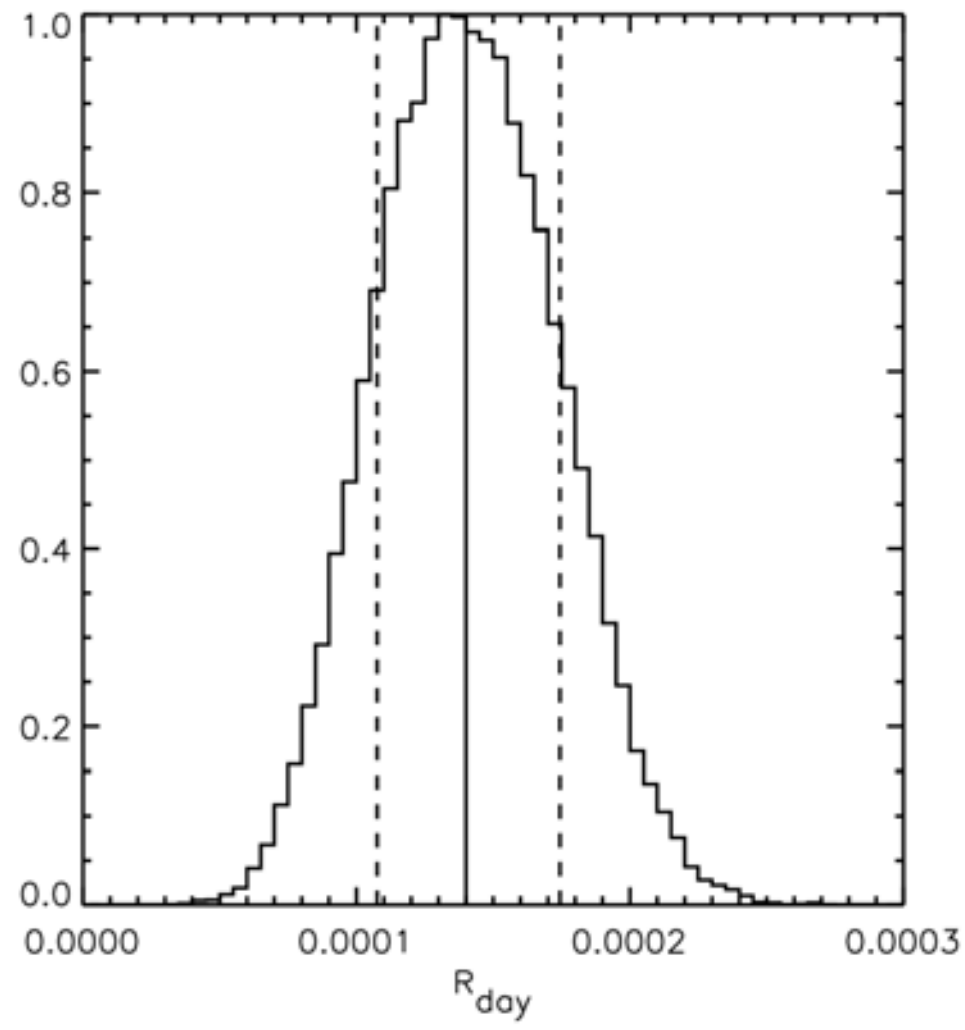

***Fig 6:*** *The probability distribution of $R_{Day}$ from the MCMC simulations. The solid line shows the median value. The dashed lines enclose 68% of the results with equal probability on either side of the median.*

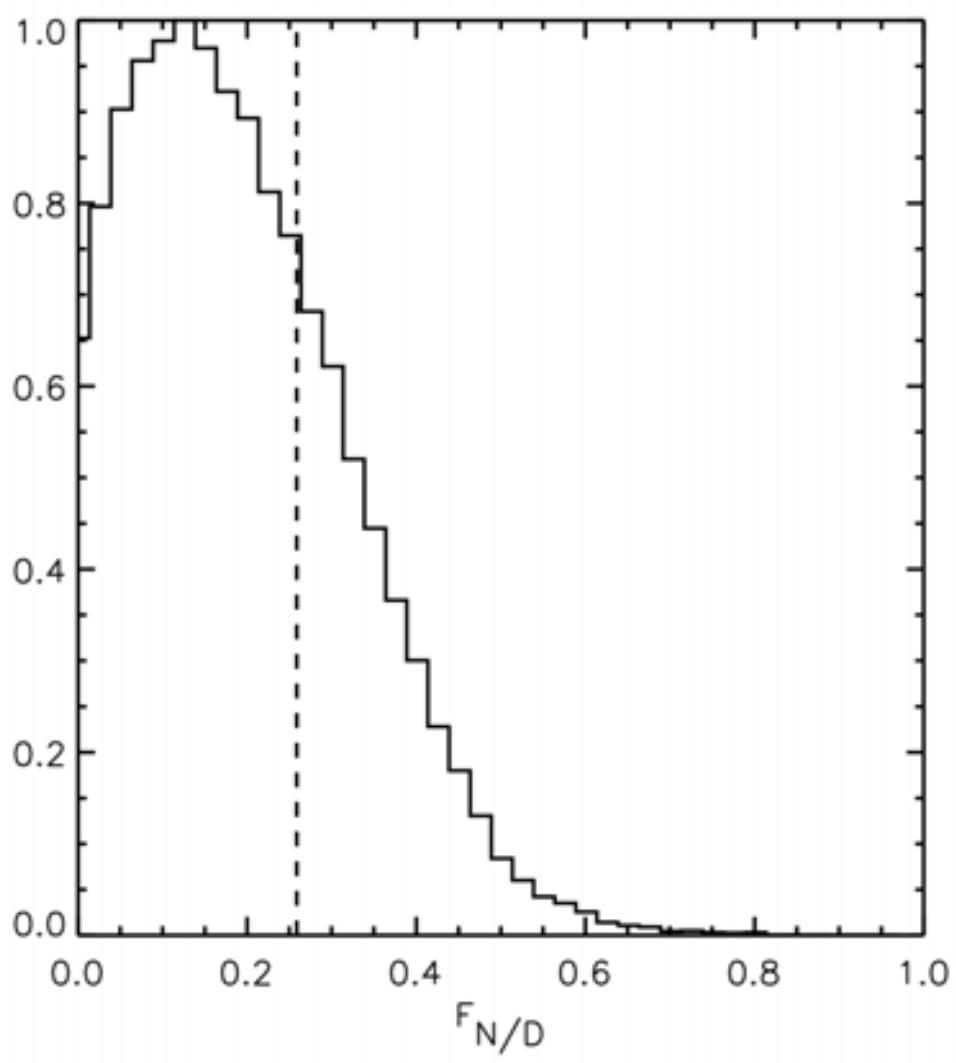

***Fig. 7***: *The probability distribution of $F_{N/D}$ as estimated from the MCMC simulations. The dashed line indicates the 68% upper limit.*

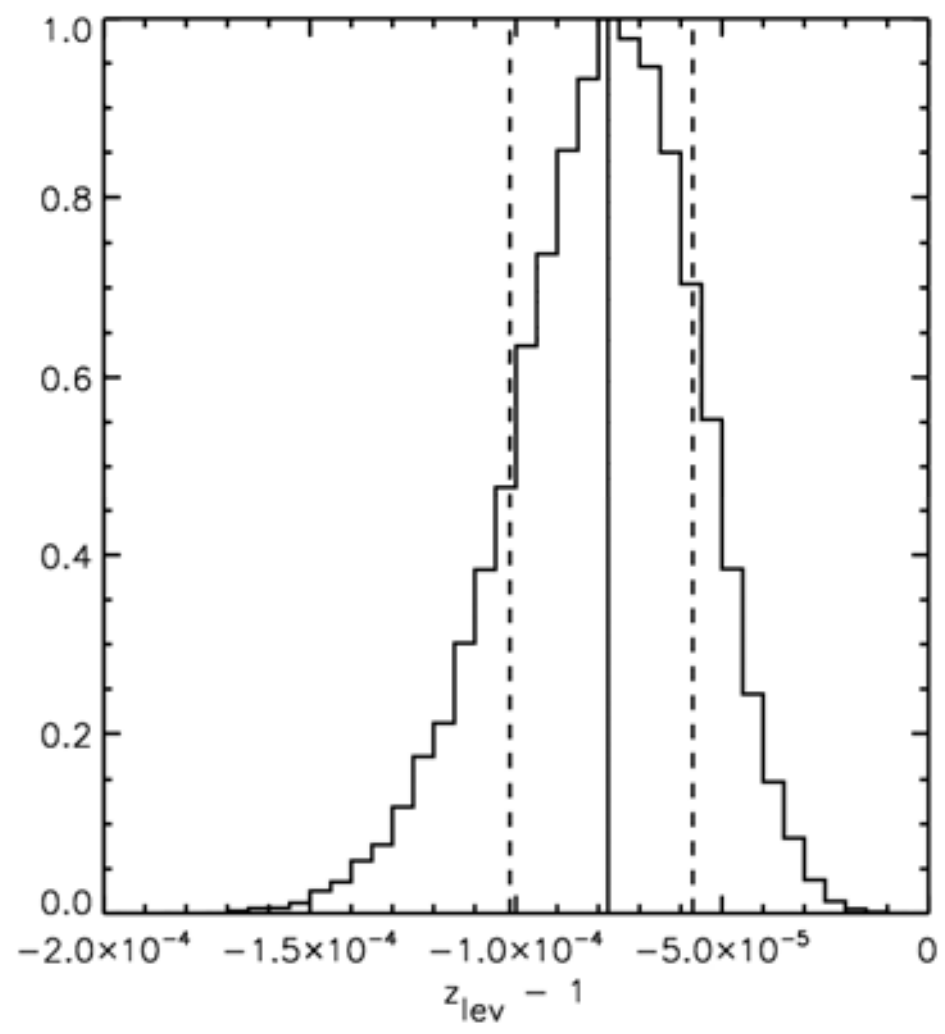

***Fig 8*** *The probability distribution of the stellar brightness expressed as fractional deviation from the mean flux in the light curve. The dashed lines enclose 68% of the results with equal probability on either side of the median.*

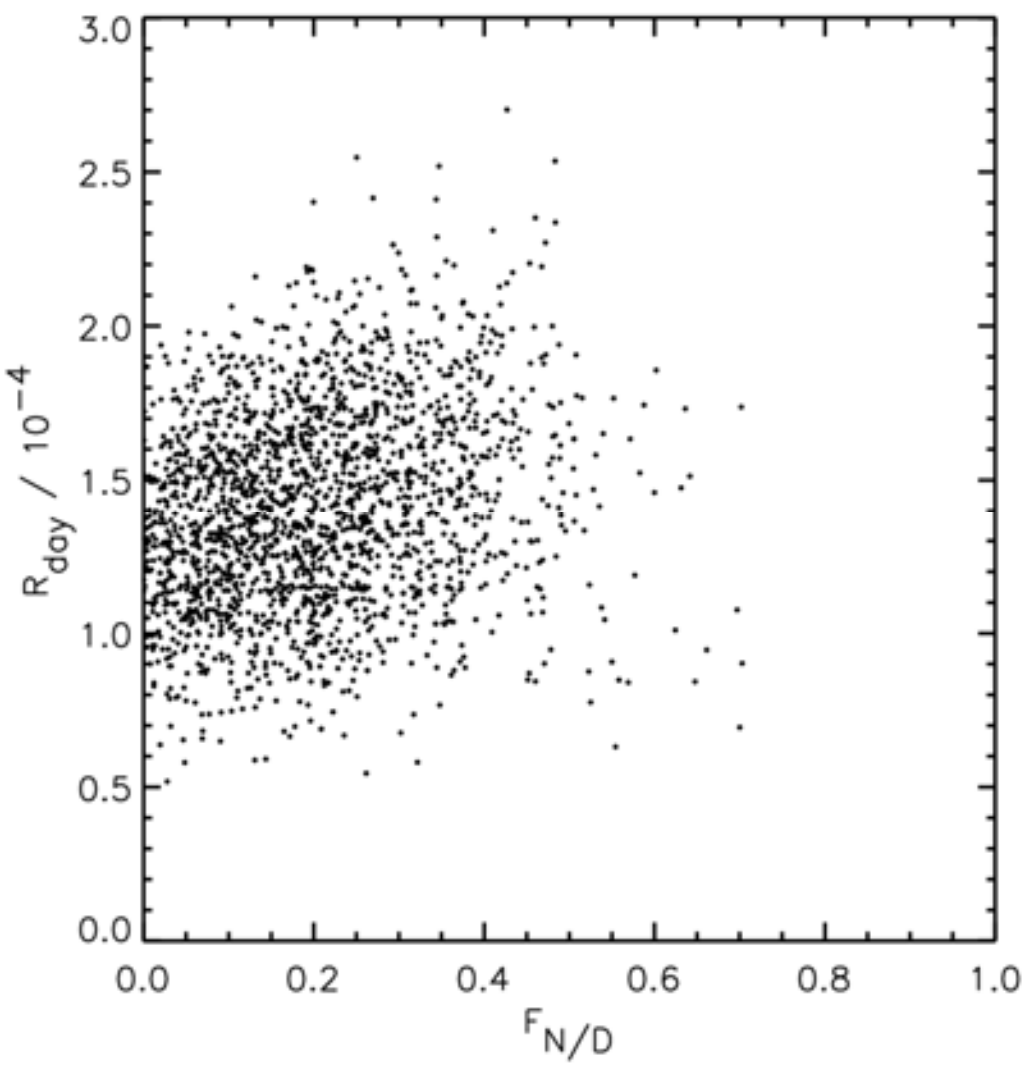

*__Fig 9:__ $R_{Day}$ versus $F_{N/D}$ from the MCMC simulation, showing a randomly selected subsample of 2x103 points for plotting purposes*

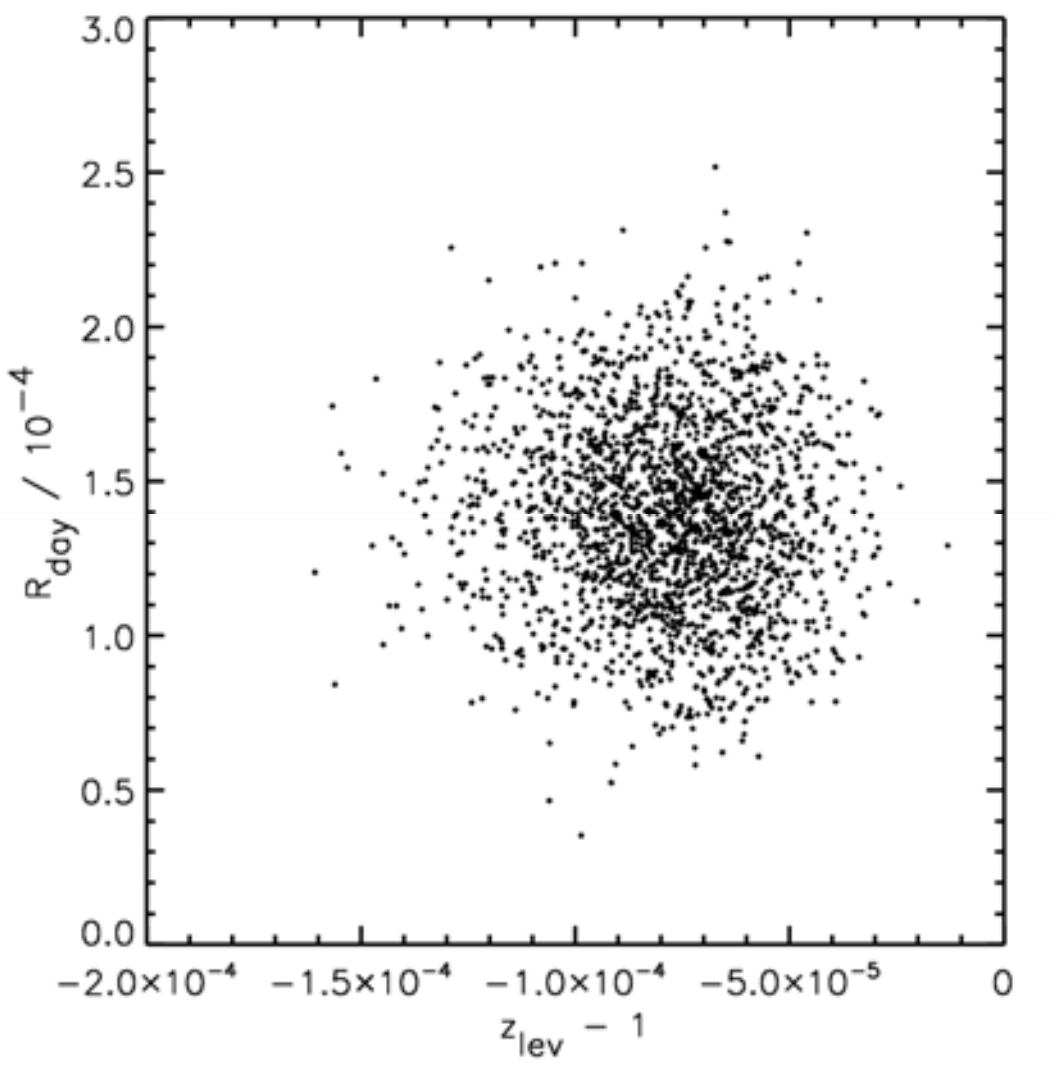

*__Fig 10:__ As in Fig. 9 for $z_{lev}$ versus $R_{Day}$.*

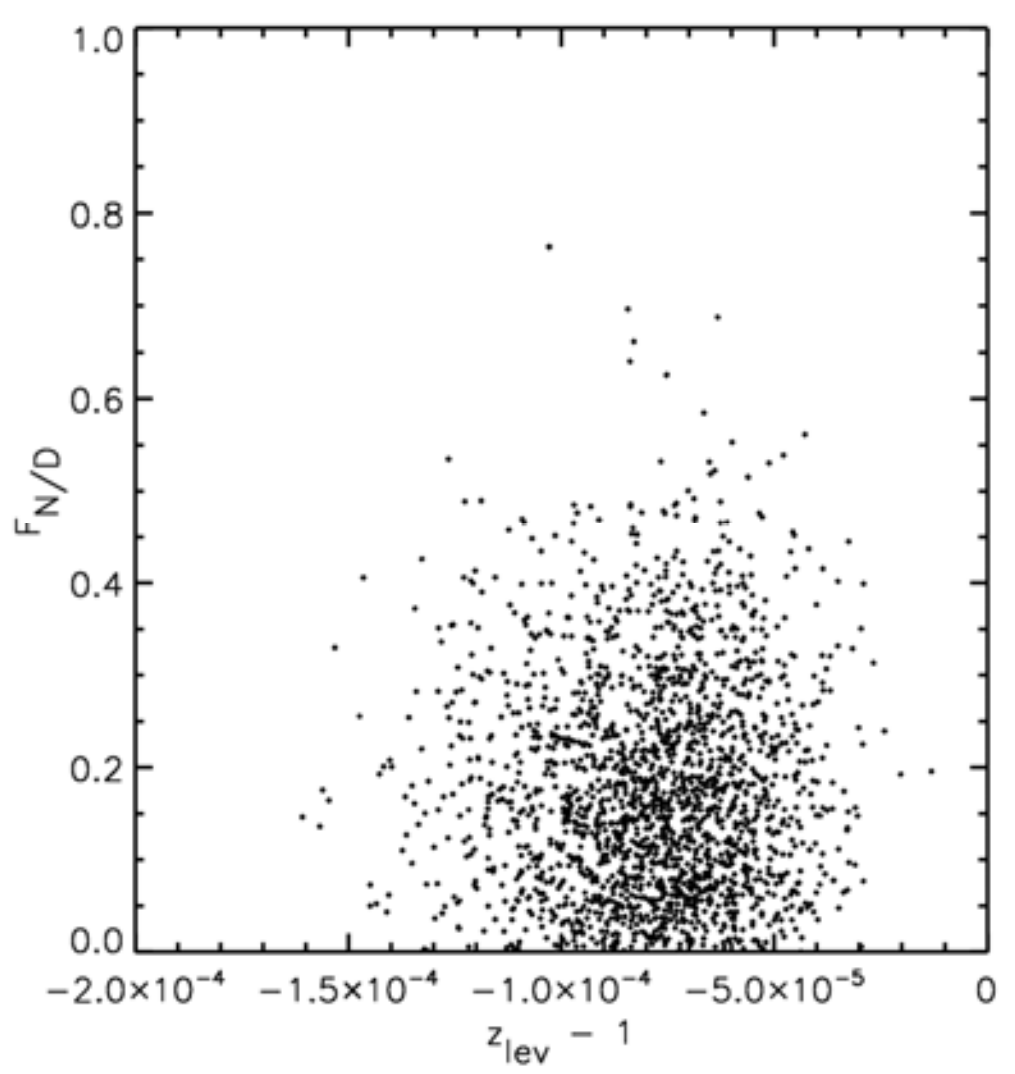

*__Fig 11:__ As in Fig. 9 for $z_{lev}$ versus $F_{N/D}$ .*